\documentclass[]{interact}

\usepackage{epstopdf}
\usepackage[caption=false]{subfig}
\usepackage{cite}

\usepackage{soul}
\theoremstyle{plain}

\usepackage{xcolor}

\theoremstyle{definition}

\theoremstyle{remark}

\begin{document}


\title{Rheology of active emulsions with  negative effective viscosity}

\author{
\name{I.Favuzzi\textsuperscript{a}, L.N. Carenza\textsuperscript{a,b}, F. Corberi\textsuperscript{c}\thanks{CONTACT F. Corberi. Email: fcorberi@unisa.it}, G. Gonnella\textsuperscript{a}, A. Lamura\textsuperscript{d} and G. Negro\textsuperscript{a}}
\affil{\textsuperscript{a} Dipartimento di Fisica, Universit{\' a} degli Studi di Bari and INFN, via Amendola 173, Bari, I-70126, Italy; \textsuperscript{b} Instituut-Lorentz, Universiteit Leiden, P.O. Box 9506, 2300 RA Leiden, Netherlands; \textsuperscript{c} Dipartimento di Fisica “E. R. Caianiello”, and INFN, Gruppo Collegato di Salerno, and CNISM, Unit{\' a} di Salerno, Universit{\' a} di Salerno, Via Giovanni Paolo II 132, 84084 Fisciano (SA), Italy;
\textsuperscript{d} Istituto Applicazioni Calcolo - CNR, Via Amendola 122/D, 70126 Bari, Italy}
}

\maketitle

\begin{abstract}
We numerically study by lattice Boltzmann simulations the rheological properties of an active emulsion made of a suspension of an active polar gel embedded in an isotropic passive background. We find that the hexatic equilibrium configuration of polar droplets is highly sensitive to both active injection and external forcing and may either lead to asymmetric unidirectional states which break top-bottom symmetry or symmetric ones. In this latter case, for large enough activity, the system develops a shear thickening regime at low shear rates. Importantly, for larger external forcing a regime with stable negative effective  viscosity is found. Moreover, at intermediate activity a region of multistability is encountered and we show that a maximum entropy production principle holds in selecting the most favorable state. 
\end{abstract}


\section{Introduction}
Active fluids are a fascinating class of soft materials which inherently evolve out of equilibrium due to the ability of their fundamental constituents to convert internal energy into motion~\cite{marchetti2013,ramaswamy2010}.
This leads to the occurrence of a plethora of unexpected behaviors which are unobserved in their equilibrium counterpart, such as spontaneous flow~\cite{kruse2004,giomi2008,simha2002,orlandini2008}, motility induced phase separation~\cite{digregorio2018,Chiarantoni_2020,cates2015}, active turbulence~\cite{dombrowski2004,wensink2012,carenza2020,carenza2020_bif} and many others~\cite{tjhung2012,Doostmohammadi2019,Carenza22065,carenza2020_physA,Negro_2019,recas2020}.

Apart from the important theoretical interest due to their intrinsic non-equilibrium behavior, active gels have gatheblack much attention in the scientific community because of their possible implementation in the design of novel smart materials and micro-devices.
These include, for instance, biological based \emph{labs-on-a-chip}~\cite{dutse2011microfluidics} --integrated devices of microscopic size which are able to perform analysis tasks through microfluidic measurements-- and micromotors~\cite{Thampie2016mm} --microscopic motors which may exploit the energy provided at small scales by active swimmers to produce autonomous and controllable movement of a larger apparatus.
Among potential applications, these devices may have a revolutionary role in non-invasive clinic investigation and specific drug-delivery, paving the way towards the development of a new generation of therapies for cancer and cardiac diseases.

To this aim, it is fundamental to understand the response of active systems to an external forcing and its effects on the rheological properties of the suspension~\cite{hatwalne2004,marenduzzo2007pre,Pagonabarraga2013}.
Recent research in this field has unveiled a number of unexpected behaviors which are strongly related to the complex interaction between the external forcing, which can be experimentally controlled, and the active one generated by the swimmers.
An active particle in a fluidic environment can be broadly classified either as \emph{extensile} or \emph{contractile}, in accordance to its swimming mechanism.
The former pushes the fluid at its ends, which is expelled along the long axis of the swimmer and drawn inward toward the center. The resulting 
far-flow field is dipolar and analogous to the one produced by an out-warding stresslet~\cite{yeomans2017}. Conversely, contractile swimmers behave as pullers and the mechanism is basically reversed.

Importantly, both experiments on active suspensions and theoretical investigations on active gel theory have proved that the swimming mechanism of the active constituents plays a very relevant role on the rheological properties of the system~\cite{PhysRevE.83.041910,Liverpool2006,foffano2012}.
 Indeed, extensile swimmers are able to  strengthen the externally imposed flow, thus inducing the lowering of the effective viscosity, while contractile swimmers develops transverse counteracting flows with the final effect that the viscosity of the suspension increases~\cite{Giomi2010}.

Interestingly, viscosity blackuction have been observed by L{\' o}pez \emph{et al.}~\cite{lopez2015} to give rise to intermittent superfluidic regimes in the case of dense bacterial suspensions of \emph{E. Coli} sheablack in a Couette rheometer, showing that the activity of pusher swimmers coupled to the external forcing is able to fully overcome viscous effects.
Further evidence of the emergence of such an odd rheological behavior have been more recently reported even for the case of a thin bacterial film under simple planar shear by Guo \emph{et al.}~\cite{Guo201722505}.
In this case, the stress in the system is found to develop heterogeneous states which may eventually lead to a superfluidic regime.

The occurrence of states flowing at null effective viscosity was first theoretically speculated by Cates \emph{et al.}~\cite{Cates2008} in a study on active gels where the development of inviscid flows was proposed as a possible solution to the appearance of a non monotonic region in the stress-strain ($\sigma-\dot{\gamma}$) characteristic, theoretically obtained. This would lead to a fluid flowing with negative viscosity, discarded as nonphysical, being intrinsically unstable.
Interestingly enough, intermittent regimes with negative viscosity were later observed in the experiment of L{\' o}pez \emph{et al.}, as a transient response to switching off the Couette rheometer.

This challenging topic was recently consideblack by Loisy \emph{et al.}~\cite{loisy2018} in a numerical study that showed 
that negative effective viscosity is due to a non-monotonic local velocity profile, in a quasi-$1d$ system using a minimal model for active liquid crystals.
Later on, the authors of this paper have addressed the shear thinning mechanism in a comprehensive bidimensional model for polar active emulsions, finding an intermittent multistable dynamics with the appearance of both inviscid and negative viscosity regimes.

However, in order to implement active systems for the design of novel devices, it is fundamental to control and trigger the onset of each rheological state. 
In this article we will show that it is actually possible to select and stabilize a particular
rheological regime by opportunely setting parameters which are experimentally controllable.
In particular, we shall here consider a system of multiple active polar droplets emulsified in a passive isotropic background sheablack between two moving walls. The system that we consider~\cite{negro2018,carenza2019_IJMP,Negro2019_proceeding,negro2019,scientificreports2020} has the property that a tunable amount of active material can be homogeneously dispersed in an emulsion. \textcolor{black}{Importantly, the two components have the same nominal viscosity, so that the observed rheological behaviors uniquely result from the complexity introduced by the mutual effect of interfaces, liquid crystalline phase and activity.
Therefore, in absence of a Newtonian background, one would obtain an uniform liquid crystalline suspension whose rheological properties are well known in literature, both for passive and active preparations ~\cite{mariagrazia,Markovich2019,Thampi2015}.}

\textcolor{black}{In the following Section we will present the dynamical model and the numerical approach, while in Section~\ref{sec:results} the observed rheological regimes will be discussed and classified. In particular,} by systematically varying both the rate of active injection and the external forcing, we will show that a series of morphological and rheological transitions takes place, resulting in the development of both negative effective viscosity states and inviscid regimes, as well as shear thickening.
\textcolor{black}{In the following Sections, each of the aforementioned regimes will be consideblack in detail. Section~\ref{sec:intermittent} will be devoted to the analysis of the onset of the activity induced instability which leads to symmetry breaking and to the consequent intermittent dynamics and we will furnish a mechanistic explanation in terms of the dynamics of the polar liquid crystal.
Finally, in Section~\ref{sec:large} we will discuss how the combined effects of shear and large activity may give rise either to shear thickening or to an effective negative viscosity.}

\section{The model}
We consider a system comprising an emulsion of active material suspended in a Newtonian fluid with mass density $\rho$.
To describe the physics of the system we make use of an extension of the well-established active gel theory~\cite{ramaswamy2010,marchetti2013,tjhung2012,tjhung2015,kruse2004}.
In this context, we will consider the density $\rho$ and the fluid velocity  $\mathbf{v}$ as hydrodynamic variables. Moreover, we introduce the concentration $\phi$ of active material and the polarization field $\mathbf{P}$ which defines the local average orientation of the active constituents.
The temporal evolution of the system is ruled by the following set of partial differential equations:
\begin{eqnarray}
\rho\left(\frac{\partial}{\partial t}+\mathbf{v}\cdot\nabla\right)\mathbf{v} & = &  \nabla\cdot \tilde{\sigma}^{tot}\ ,\label{nav}\\
\frac{\partial \phi}{\partial t}+\nabla\cdot\left(\phi\mathbf{v}\right)&=& M \nabla^2  \mu,\label{conc_eq}\\
\frac{\partial\mathbf{P}}{\partial t}+\left(\mathbf{v}\cdot\nabla\right)\mathbf{P}&=&-\tilde{\Omega}\cdot\mathbf{P}+\xi\tilde{D}\cdot\mathbf{P}
-\frac{1}{\Gamma} \mathbf{h}.\label{P_eq}
\end{eqnarray}
The first is the incompressible Navier-Stokes equation, where $\tilde{\sigma}^{tot}$ is the total stress tensor \cite{beris1994}. This can be divided into an equilibrium/passive and a non-equilibrium/active part:
\begin{equation}
\tilde{\sigma}^{tot}=\tilde{\sigma}^{pass}+\tilde{\sigma}^{act}.
\end{equation}
The passive contribution $\tilde{\sigma}^{pass}$ takes into account the viscous dissipation as well as
the elastic response of the liquid crystal and the binary fluid.
The passive term is in turn the sum of four contributions:
\begin{equation}
\tilde{\sigma}^{\textit{pass}}=\tilde{\sigma}^{\textit{hydro}}+\tilde{\sigma}^{\textit{visc}}+\tilde{\sigma}^{\textit{pol}}+\tilde{\sigma}^{bm} \mbox{.}
\label{eqn:passive_stress_tensor}
\end{equation}
The first term is the hydrodynamic pressure contribution given by $\sigma^{\textit{hydro}}_{\alpha\beta}=-p\delta_{\alpha\beta}$. The incompressible expression for the viscous stress is given by $\sigma_{\alpha\beta}^{visc} = \eta_0(\partial_{\alpha}v_{\beta} +\partial_{\beta}v_{\alpha})$, where $\eta_0$ is the nominal viscosity of the fluid. The polar elastic stress is analogous to the one used in nematic liquid crystals~\cite{beris1994}:
\begin{eqnarray}
\sigma_{\alpha\beta}^{pol}=\frac{1}{2}(P_{\alpha}h_{\beta} -P_{\beta}h_{\alpha})-\frac{\xi}{2}(P_{\alpha}h_{\beta}+P_{\beta}h_{\alpha})\nonumber\\
- k_{P} \partial_{\alpha}P_{\gamma}\partial_{\beta}P_{\gamma}\label{eq:elastic-stress},
\end{eqnarray}
where 
$\xi$ is a constant controlling the aspect ratio of active particles (positive for rod-like particles and negative for disk-like ones),
$k_P$ is the liquid crystal elastic constant and $\mathbf{h}=\delta \mathcal{F}/\delta\mathbf{P}$ is the molecular field with $\mathcal{F}$ a suitable free energy to be defined in the following. 
The magnitude of $\xi$ also determines the response of the liquid crystal to an external shear flow: For $|\xi|>1$ particles align to the imposed flow (flow aligning regime), while for $|\xi|<1$ the resulting dynamics is never stationary and the polarization field rotates in the direction defined by the imposed shear flow (flow tumbling regime).
The last term on the right-hand side of Eq.~\eqref{eqn:passive_stress_tensor} accounts for interfacial stress:
\begin{equation}
\sigma_{\alpha\beta}^{bm}=\left( f-\phi\frac{\delta F}{\delta\phi} \right)\delta_{\alpha\beta} - \frac{\delta F}{\partial\left(\partial_{\beta}\phi\right)} \partial_{\alpha}\phi,
\end{equation}
where $f$ denotes the free energy density.
Finally, the active stress tensor has phenomenological origin and does not stem from the free energy. Its expression in terms of the order parameters is
\begin{equation}
\sigma_{\alpha\beta}^{act}=-\zeta \phi \left(P_{\alpha}P_{\beta}-\frac{1}{3}|{\bf P}|^2\delta_{\alpha\beta}\right)\label{eq:active-stress}
\end{equation}
and can be obtained by coarse-graining over an ensamble of force dipoles~\cite{simha2002}. 
Here $\zeta$ is the activity parameter. This is positive for extensile systems (pushers) and negative for contractile ones (pullers). The active stress drives the system out of equilibrium by injecting energy on the typical length-scales of deformation of the polarization pattern.

Eqs.~\eqref{conc_eq} and \eqref{P_eq} respectively define the time evolution of the concentration of the active material and of
the polarization field. 
In particular the former is a convection-diffusion equation, based on the assumption that the concentration field is locally conserved. Here $M$ is the mobility and $\mu=\delta F/\delta\phi$ the chemical potential. The polarization field follows an advection-relaxation equation, Eq.~\eqref{P_eq}, borrowed from polar liquid crystal theory. $\Gamma$ is the rotational viscosity, while
$\tilde{D}=(\tilde{W}+\tilde{W}^T)/2$ and $\tilde{\Omega}=(\tilde{W}-\tilde{W}^T)/2$ stand for the symmetric and the anti-symmetric parts of the velocity gradient tensor $W_{\alpha\beta}=\partial_{\beta}v_{\alpha}$. 

The equilibrium properties of the system in absence of activity are defined by the following free-energy functional based on the Brazovskii theory~\cite{braz1975,gonnella1997,Corberi1999} for weak crystallization, extended for the treatment of a polar liquid crystal~\cite{bonelli2019,negro2018}:
\begin{eqnarray}\label{eqn:fe}
&F[\phi,\mathbf{P}]
=\int d\mathbf{r}\,\{\frac{a}{4\phi_{cr}^4}\phi^{2}(\phi-\phi_0)^2+\frac{k_\phi}{2}\left|\nabla \phi\right|^{2}+\frac{c}{2}(\nabla^2\phi)^2 \nonumber\\
&-\frac{\alpha}{2} \frac{(\phi-\phi_{cr})}{\phi_{cr}}\left|\mathbf{P}\right|^2+ \frac{\alpha}{4}\left|\mathbf{P}\right|^{4}+\frac{k_P}{2}(\nabla\mathbf{P})^{2}
+\beta\mathbf{P}\cdot\nabla\phi\} \ \ .
\end{eqnarray}
For $a>0$, the concentration field has two minima at $\phi=0,\phi_0$.
The second and third terms determine the surface tension of the system. In particular, by allowing $k_\phi$ to become negative, formation of interface becomes energetically favoblack while $c$ has to be positive for thermodynamic stability~\cite{braz1975}. 
The polynomial terms in $|\mathbf{P}|$, where $\alpha$ is positive, 
allow for the segregation of the polarization field in those regions where  $\phi > \phi_{cr}$, being $\phi_{cr}$ a reference value which allows us to discriminate passive (isotropic) regions ($\phi < \phi_{cr}$) from the active/polar ones.
The term proportional to the polarization gradient pays the energetic cost for liquid crystal deformations.
Finally, the last term defines the anchoring properties of the polarization field at the interface. Homeotropic anchoring is achieved by setting $\beta \ne 0$. In this case, the polarization field either points towards the passive  phase if $\beta>0$ or the active one otherwise.

For symmetric compositions of the system --where the two components are equally represented-- a transition from the ordeblack phase towards the lamellar phase is found at $a=k^2_\phi/4c + \beta^2/k_P$, with lamellar width given by the Brazovskii length-scale  $\lambda=2 \pi \sqrt{2c/|k_\phi|}$.
However, for enough asymmetric compositions ($\phi_{cr} \lesssim 0.35$ with the bar denoting space average) the system sets into an emulsion of polar droplets suspended in an isotropic bacgkround and arranged in a hexatic pattern~\cite{Henrich2012} (see for instance panel~(a) of Fig.~\ref{fig2}).

\textcolor{black}{
The other relevant scales of the theory are: the coherence length of the polar liquid crystal $l_P=\sqrt{k_P/\alpha}$ which controls how quickly the order parameter drops in
the neighborhood of a topological defect; and  the active length-scale $l_a=\sqrt{k_P/|\zeta|}$~\cite{giomi2014_2} which defines the typical  scale of elastic deformations due to active injection.
In particular the model parameters are chosen to have $\lambda \ll L$ ($L$ being the system size) and $l_P \sim \mathcal{O}(1) < \lambda $ in order to guarantee enough resolution of the liquid crystal pattern. Finally, by varying the activity parameter $\zeta$ we are able to move the system from a passive state, where active injection does not alter significantly the dynamics, to a proper \emph{active} state. These two regimes correspond to situations with $\lambda < l_a$  and    $l_a < \lambda$, respectively. In this latter case the liquid crystal undergoes activity-induced elastic instabilities.}
Therefore, the model here presented provides an effectful and easy way to confine the active material on a well defined scale, thus allowing for the direct control of the typical length-scale at which energy is injected in the system.

\textcolor{black}{The adimensional numbers controlling the system are the Ericksen number $Er=\eta_0 \dot{\gamma}/B$, with $B=(\beta^2/k_P + k_\phi^2/c)\phi_0^2$ the compression modulus~\cite{negro2019}, that is often used in the study of liquid crystals to describe the deformation of the orientational order parameter field under flow, and the active Ericksen number $Er_{act}=\zeta/B$~\cite{giomi2014_2}. However, in the following we will present our results in terms of the activity parameter $\zeta$ and shear rate $\dot\gamma$.}

\subsection{Numerical Method and Parameters}

Eqs.~\eqref{nav}-\eqref{P_eq} have been solved numerically by means of a well validated hybrid lattice Boltzmann (LB) approach (in the limit of incompressible flow). More in detail, the Navier-Stokes equation was solved through a pblackictor-corrector LB scheme~\cite{carenza2019}, while the evolution equations for the order parameters $\phi$ and $\mathbf{P}$ were integrated through a pblackictor-corrector finite-difference algorithm
implementing first-order upwind scheme and fourth-order accurate stencils for space derivatives.
We made use of a parallel approach implementing Message Passage Interface (MPI) to parallelize the code through the \emph{ghost-cell} approach~\cite{MPI}.

Simulations were performed on a $2d$ square lattice (D2Q9) of size $L=256$. The system was initialized in a mixed state, with $\phi$ uniformly distributed between $1.1$ and $0.9$ being $\phi_{cr}=1$. 
The concentration $\phi$ ranges from $\phi=0$ (passive phase) to $\phi\simeq 2$ (active phase). 
Unless otherwise stated, parameter values are $a=4\times 10^{-3}$, $k_\phi=-6\times 10^{-3}$, $c=10^{-2}$, $\alpha=10^{-3}$, $k_P=10^{-2}$, $\Gamma=1$, $\xi=1.1$, $\phi_0=2$, $\beta=10^{-2}$, $\eta_0=1.67$.

The system is confined in a channel with no-slip boundary conditions at the bottom and top walls located at $z=0$ and $z=L$, respectively 
(the $z$-axis is the shear direction), 
implemented by bounce-back boundary conditions for the distribution functions~\cite{succi2001} in the LB algorithm.
Periodic boundary conditions were imposed along the (flow) $y$-direction. The shear flow was imposed by moving walls in opposite directions, respectively with velocity $v_w$ for the top wall and $-v_w$ for the bottom wall, so that the imposed
shear rate is given by $\dot{\gamma}=\frac{2 v_w}{L}$.

Neutral wetting boundary conditions were enforced by requiring on the wall sites 
\begin{align}
\nabla_{\perp} \mu \rvert_\textrm{walls} =  0 \ , \ \ \  
\nabla_{\perp} (\nabla^2 \phi) \rvert_\textrm{walls} =  0 \ ,
\end{align}
where $\nabla_{\perp}$ denotes the partial derivative computed normally to the walls and directed towards the bulk of the system.
Here the first condition ensures density conservation and the second determines the wetting to be neutral. 
Moreover, strong tangential anchoring was imposed for $\mathbf{P}$ on the walls:
\begin{equation}
P_\perp \rvert_\textrm{walls} =  0, \qquad \nabla_\perp P_\parallel \rvert_\textrm{walls}=  0 ,
\end{equation}
where $P_\perp$ and $P_\parallel$ denote, respectively, normal and tangential components of the polarization field with respect to the walls.

\section{Numerical Results}
\label{sec:results}
\begin{figure}[t]
	\centerline{\includegraphics[width=0.98\columnwidth]{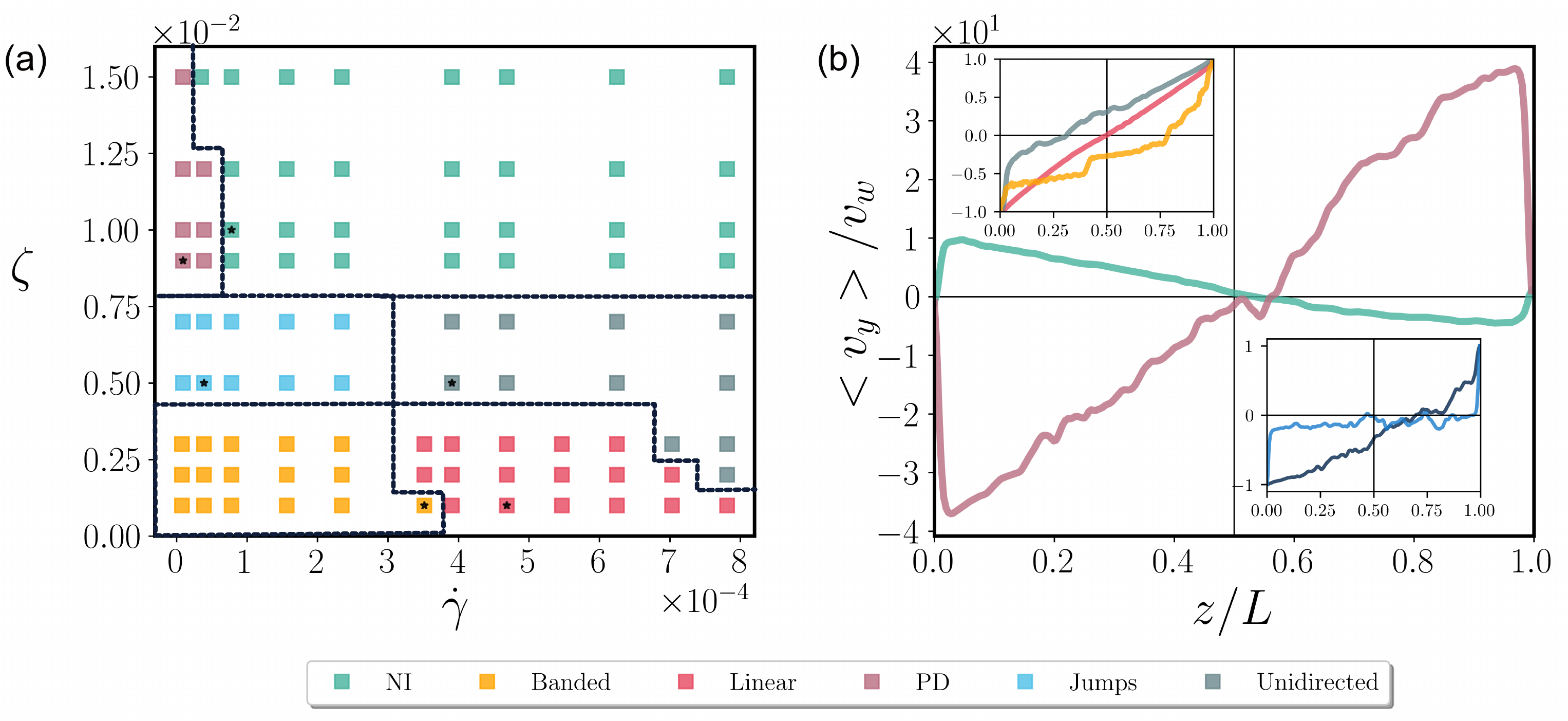}}
	\caption{\textbf{Rheological regimes and velocity profiles.} Panel (a) shows the different rheological regimes encounteblack at 
varying both the activity $\zeta$ and the shear rate $\dot{\gamma}$. Black stars denote  the cases shown in panel (b) and in the following figures.
Panel (b) shows the corresponding velocity profiles across the
channel after averaging the component $v_y$ along the flow direction. 
At very low activity 
$\zeta<\zeta_{cr}$ (black and yellow symbols) the system is found in a passive regime where active injection does not sensibly affect the response
 of the system. In this case, the velocity profile  is either linear (black line in the upper inset of  panel~(b)) at large shear rates or 
exhibits shear bandings at low $\dot{\gamma}$ (yellow line in the upper inset of panel~(b)). As activity is increased the dynamics is 
characterized by the breaking of the top-bottom symmetry for $\dot{\gamma} \geq 4 \times 10^{-4}$ (gray symbols). The system now preferentially 
flows in just one direction (see gray line in the upper inset of panel~(b)) with the position of the inversion region ($\langle v_y \rangle=0$) 
closer to one of the two walls. In the same range of activity and shear rate $\dot{\gamma}<4 \times 10^{-4}$ an intermittent dynamics is 
observed (blue symbols) with the system \emph{jumping} between different rheological states in a random fashion (see blue and light blue lines
in the lower inset of panel~(b)). 
Finally, as activity is 
increased over $\zeta>0.8 \times 10^{-2}$ two stable regimes possibly occur. For $\dot{\gamma}<1 \times 10^{-4}$ the system exhibits shear 
thickening (purple symbols) characterized by the enhancement of the imposed velocity in a thin layer in proximity of the two walls 
(see purple line in panel~(b)). At larger activity, a state flowing at negative effective viscosity is observed (green symbols). In this case, 
the effect of activity is to produce the inversion of the flow in proximity of the two layers, with the velocity profile in the channel bulk 
flowing with an inverted slope (green line in panel~(b)). The color code used to plot the velocity profiles in panel~(b) matches the color 
of the corresponding region in the phase diagram of panel~(a). }\label{fig1}
\end{figure}

We start by discussing the rheological behaviors encounteblack at varying systematically both the shear rate $\dot{\gamma}$ and the activity $\zeta$
which will be always consideblack positive in the present study.
Fig.~\ref{fig1} shows the phase diagram in the $\zeta-\dot{\gamma}$ plane (panel (a)) and the typical velocity profiles \textcolor{black}{at steady-state} (normalized by the wall velocity $v_w$) across the channel, associated to each rheological regime (panel (b)).

\paragraph*{\textcolor{black}{The passive limit.}} Before getting involved into the description of the activity-induced effects, we shall comment on the passive limit, corresponding to a situation where activity is either null or smaller than a critical value $\zeta_{cr}(\dot{\gamma})$. \textcolor{black}{This threshold depends on the intensity of the shear rate $\dot{\gamma}$ and it is chosen by comparing the rheological and morphological state with the one of the corresponding reference state at $\zeta=0$ and same $\dot{\gamma}$. Therefore, in the passive limit the dynamics of the system  is not influenced by active energy injection --a regime that we shall refer to as \emph{quiescent}.
As shown in Fig.~\ref{fig1} it results to be $\zeta_{cr} \simeq 4.0 \times 10^{-3}$ ($Er_{act}=0.073$) for small shear rates while it
decreases down to $\zeta_{cr} \simeq 2.5 \times 10^{-3}$ ($Er_{act}=0.045$) for $\dot\gamma \gtrsim 6 \times 10^{-4}$ ($Er \gtrsim  0.0018$).}
In this case, for high shear rates ($\dot{\gamma}\gtrsim 4 \times 10^{-4}$) the velocity profile is linear (black line in the upper inset of  Fig.~\ref{fig1}(b)).
However, by blackucing the intensity of the external forcing, the velocity profiles progressively loose their linear features and one or more shear bandings develop in the channel (yellow line in the upper inset of Fig.~\ref{fig1}(b)). This behavior is related to the presence of topological dislocations in the hexatic arrangement of droplets, as we will discuss in more detail in the next Section. 
\textcolor{black}{The transition from the shear banding to the linear regime takes place at $\dot{\gamma} \simeq 3.8\times 10^{-4}$. In the following we shall address as weak shear rates those values of $\dot{\gamma}$ below such threshold and as large shear rates those values beyond it.}

\paragraph*{\textcolor{black}{Shear thickening and negative viscosity.}} By increasing activity over the critical threshold $\zeta_{cr}$, the energy injected in the system by the active component drastically influences both the morphological and the rheological state and a plethora of unexpected behaviors appears, including stable unidirected profiles (gray line in the upper inset of Fig.~\ref{fig1}(b)), inverted profiles denoted as NI (green line) and enhanced profiles denoted as PD (purple line).
The nomenclature that we use to identify these regimes refers to the corresponding rheological state: The first letter, either 
P or N, refers to the sign of the measublack viscosity (either positive or negative) while the second letter, either  D or I, 
corresponds to the fact that velocity profiles may either be directed or inverted.
For instance, in the case of enhanced profiles (PD) at large activity ($\zeta>0.8 \times 10^{-2}$) the fluid is boosted in the same direction as the wall velocity so that the slope of the observed profile (purple line in Fig.~\ref{fig1}(b)) has the same sign of the imposed one, 
but it is steeper. This results in the increment of the effective viscosity, since the stress in the bulk is larger than the viscous contribution.
Conversely, inverted profiles (NI, green line) are characterized by an opposite behavior. The intensity of the flow is drastically blackuced in proximity of the walls and eventually it gets inverted, so that the fluid in the bulk of the channel flows in the opposite direction with respect to the imposed one. This results from the fact that active pumping at the boundaries opposes to the external forcing. Moreover, the active shear stress is larger (in modulus) than reactive and viscous contributions, resulting in a state which flows with effective negative viscosity.

\paragraph*{\textcolor{black}{Intermittent dynamics and unidirected motion.}} Importantly, at intermediate values of activity and large shear rate ($\zeta_{cr} < \zeta < 7.5 \times 10^{-3} $ and $\dot{\gamma}> 4.0 \times 10^{-4}$) yet another behavior is observed. In this case, velocity blackuction is observed at  one of the two layers, while in the rest of the system  profiles are linear. The resulting state is no more symmetric and the system basically flows  in just one direction, since the inversion region moves from the center of the channel towards one of the two walls.
Moreover, by blackucing the intensity of shear rate under $\dot{\gamma}<4.0 \times 10^{-4}$ in the same range of activity, a region of multistability between different rheological states is encounteblack. In this case, the system undergoes an intermittent dynamics characterized by \emph{jumps} from the unidirectional state (blue line in the inset of Fig.~\ref{fig1}(b)) -- to a superfluidic state flowing at almost null effective viscosity with the velocity profile  (light blue line in the inset of Fig.~\ref{fig1}(b)) 
undergoing a drastic blackuction at both walls and exhibiting a flat region in the bulk.

This concludes the description of the observed rheological behaviors. In the following Section we will provide a more detailed description of the linear and unidirected regimes in terms of the dynamics of the concentration and polarization fields.

\begin{figure}[t]
	\centerline{\includegraphics[width=1.0\columnwidth]{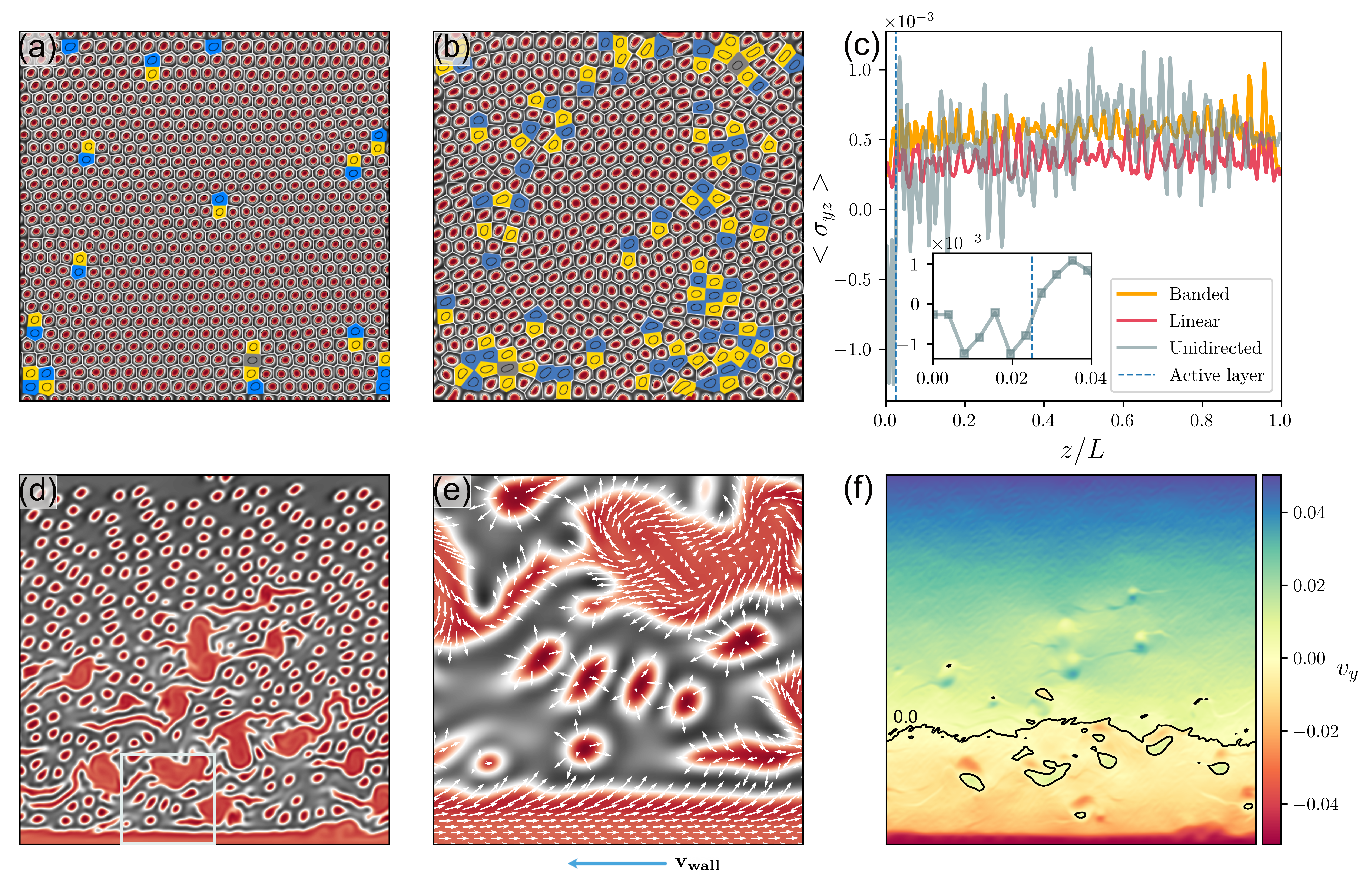}}
	\caption{\textbf{Low activity regime and unidirectional motion.} Panels~(a) and~(b) show the color plot of the 
concentration field $\phi$ (black regions correspond to active ones) at $\zeta=1\times 10^{-3}$
	and $\dot{\gamma}=3.5\times 10^{-4},4.6\times 10^{-4}$, respectively. The former is characterized by the hexatic arrangement 
of the droplets, with a few persistent dislocations which give rise to banded velocity profiles, as can be seen from the superimposed 
Voronoi tessellation in panel~(a)  (droplets with $5$ neighbors are highlighted in yellow, those with $7$ in blue, while droplets with 
$8$ neighbors  in gray). The latter case refers to the linear regime where droplets are deformed under the effect of the imposed flow 
and arrange in \textcolor{black}{a disordeblack pattern characterized by proliferation of topological defects}. Panel~(d) shows a typical unidirected configuration at $\zeta=5 \times 10^{-3}$ 
and $\dot{\gamma}=4 \times 10^{-4}$ with large rotating domains in the bulk and a thick active layer adhering on one of the wall. 
Panel~(e) shows a zoom over the region highlighted by the white frame in panel~(d) with superimposed polarization field. The flow state 
is shown in panel~(f) where the color plot of the velocity field is plotted. Here the black line denotes the inversion region where $v_y$ 
becomes null. Notice that active droplets, here characterized by a dipolar structure, rotating in the opposite direction with respect to 
the adjacent fluid. Panel~(c) shows the stress profiles averaged along the channel for the three configurations in panels~(a), (b), (d). 
The dashed line denotes the extension of the active layer from the bottom wall. 
The inset in panel (c) shows an enlargement of the stress in the unidirected cases in the bottom active layer.}\label{fig2}
\end{figure}

\section{Hexatic order, symmetry breaking and intermittent dynamics}
\label{sec:intermittent}
\begin{figure}[t]
	\centerline{\includegraphics[width=0.7\columnwidth]{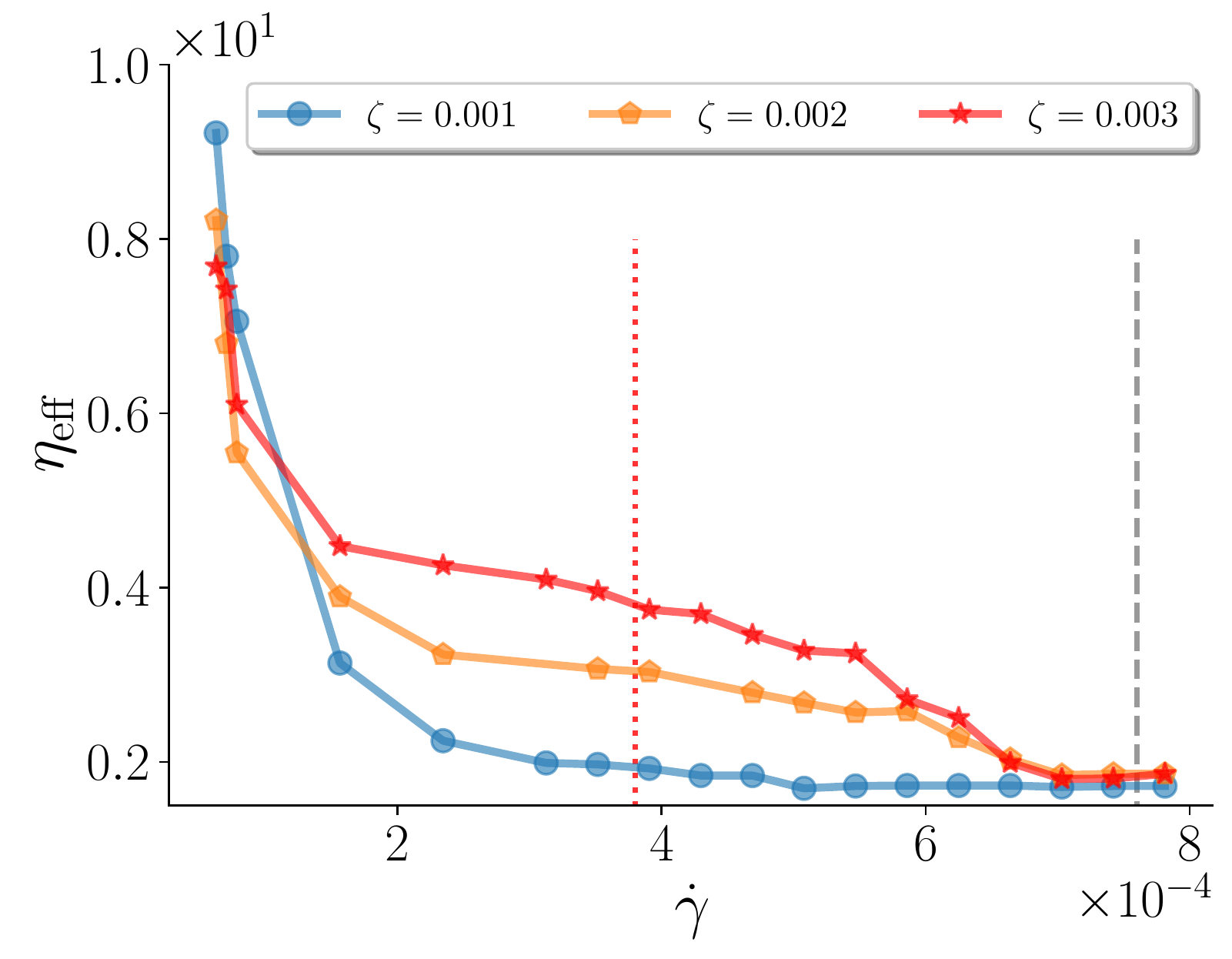}}
	\caption{\textbf{Rheological curves  in the low activity regime.} The effective viscosity $\eta_{eff}$ is plotted versus the shear rate $\dot{\gamma}$. Notice that shear thinning is observed at varying $\dot{\gamma}$, while the curves monotonically increase with activity --a behavior commonly addressed as active shear thickening. Vertical dotted (black) line and dashed (grey) line  denote the values of shear rate $\dot{\gamma}$ at which linear and unidirected regimes are first encounteblack, respectively.}\label{fig1bis}
\end{figure}

In the limit of weak activity and shear rate, the system sets into an emulsion of polar droplets suspended in a passive isotropic background (see panel~(a) of Fig.~\ref{fig2}). The configuration is ordeblack in a hexatic fashion with a few dislocations in the arrangement, as visible in the Voronoi tessellation plotted in panel~(a) of Fig.~\ref{fig2} (droplets with $5$ neighbors are highlighted in yellow, those with $7$ in blue, while droplets with $8$ neighbors  in gray).
Dislocations have a paramount effect on the rheological response of the system since they act as a source of stress which eventually determines the occurrence of permanent shear bandings even in the long term dynamics.
By increasing the shear rate and keeping the activity fixed, the imposed flow is able to deform the droplets which loose their spherical shape. \textcolor{black}{This, in turn, leads to the loss of long-ranged hexatic order with the configuration characterized by some ordeblack domains interrupted by regions rich of topological defects where droplets can easily flow with respect to each other (see panel (b) 
of Fig.~\ref{fig2}) since they are not caged anymore in a periodic lattice.}
A direct comparison between the stress profiles averaged along the channel for the banded and linear cases 
(respectively yellow and black lines of panel~(c)  of Fig.~\ref{fig2}) shows that the morphological transition from the hexatic (banded) to the
\textcolor{black}{disordeblack} (linear) configuration is accompanied by a drastic blackuction in the intensity of the shear stress.
\textcolor{black}{This is also confirmed by the behavior of the rheological curves $\eta_{eff}-\dot{\gamma}$ shown in Fig.~\ref{fig1bis}. Shear thinning is observed at increasing the shear rate, while the curves monotonically increase with the activity $\zeta$. Interestingly, at very low activity ($\zeta \leqslant	 10^{-3}$) the system exhibits a Newtonian behavior in the linear region of the phase diagram ($\dot{\gamma} \geqslant	3.8 \times 10^{-4}$), while at larger values of $\zeta$ this is only achieved for $\dot{\gamma} > 7 \times 10^{-4}$.}

Increasing the activity over the critical threshold $\zeta_{cr}$ 
greatly affects the morphological properties of the system. Droplets in the bulk begin to merge with each other generating large active domains (see panel~(d) of Fig. 2). Herein, the polarization field, which is homeotropically anchoblack to the interfaces, develops vortical  structures which rotate under the fueling effect of active injection (see panel~(e) of Fig. 2).
This behavior has been previously observed in this system in unconfined geometries and is compatible with the bending instability of extensile polar gels~\cite{kruse2004}.

A further important feature is represented by the development of a thick active layer at the moving boundaries to which the polarization field is tangentially anchoblack.
Within such layer, the active shear stress $\sigma_{yz}^{act} \sim \frac{\zeta}{2}\phi_0 P^2 \sin{2\theta}$, where $\theta$ stands for the orientation of polarization with respect to the imposed velocity ($ 0\le \theta \le \pi$). This generates an active force $f^{act}_\parallel= \partial_\perp \sigma_{yz}^{\textrm{act}}$, where $\partial_\perp$ denotes derivative in the direction normal to the walls \cite{negro2019}. 
The effect on the flowing state depends on the orientation of the polarization: On one hand, if $\mathbf{P}$ is oriented as the velocity at the wall, the imposed flow is reinforced  (since $\partial_\perp \sin 2\theta > 0$), one the other hand, if $\mathbf{P}$ is oppositely directed, this leads to a blackuction of the fluid velocity (since $\partial_\perp \sin 2\theta < 0$). 

The features here described are at the base of the observed rheological behaviors and are valid for any regime where activity is larger than the critical threshold $\zeta_{cr}$.
In particular, we shall now consider the outcome in the case of the unidirected regime. Panel~(d) of Fig.~\ref{fig2} shows the contour plot of the concentration field $\phi$ at $\zeta=5 \times 10^{-3}$ and $\dot{\gamma}=4 \times 10^{-4}$ in the gray region of the phase diagram in Fig.~\ref{fig1}(a). This is characterized by the breaking of the top bottom symmetry as the layer of active material
only develops at the bottom wall. 
Importantly, polarization in the layer is oppositely directed with respect to the imposed velocity (see panel~(e) of Fig.~\ref{fig2}). 
\textcolor{black}{This feature is reflected by the stress profile plotted in panel~(c) (continuous gray line) which attains negative values in proximity of the bottom wall. This is due to the active contribution that opposes to the external forcing and leads to an enlargement of the stress at the boundary (see Fig.~\ref{fig2}(c)), while it largely fluctuates in the bulk due to the presence of rotating domains.}
The resulting flow state is therefore asymmetric with the system mostly flowing rightwards 
as signaled by the inversion region (black line in panel~(f) of Fig.~\ref{fig2}) deep in the lower half of the channel.

\begin{figure}[t]
	\centerline{\includegraphics[width=1.0\columnwidth]{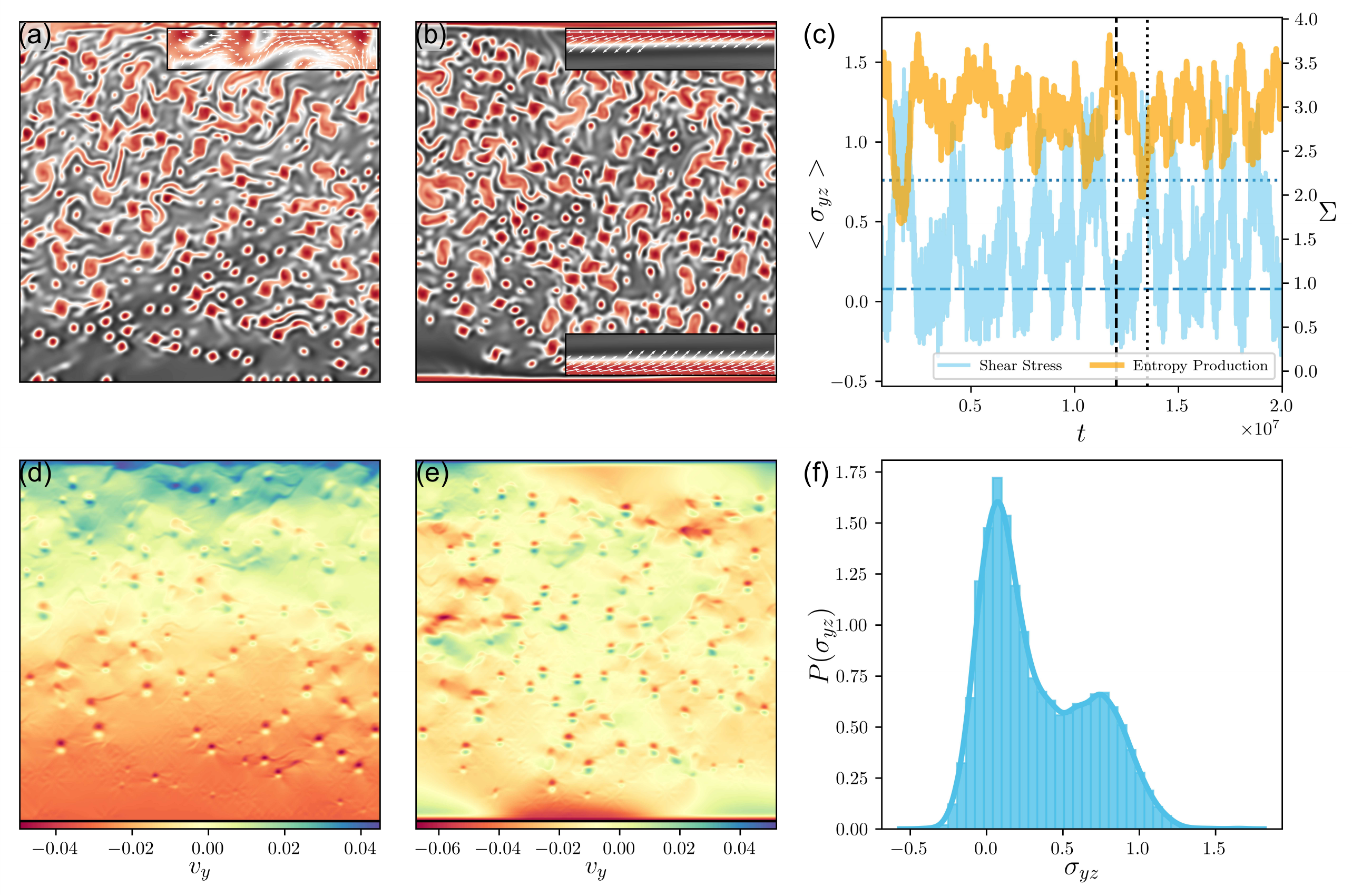}}
	\caption{\textbf{Intermittent dynamics.} Panels~(a) and~(b) show the color plot of the concentration field $\phi$ 
at $\zeta=5\times 10^{-3}$
	and $\dot{\gamma}=3.9 \times 10 ^{-5}$ for different simulation times. The configuration in~(a) ($t=1.2\times 10^{7}$) 
exhibits unidirectional flow while the one in panel~(b) is in the superfluidic state ($t=1.35\times 10^{-7}$). In both panels, 
insets show a zoom over the active layers adhering to the walls, where the white vectors define the orientation of the polarization 
field $\mathbf{P}$. Panels~(d) and (e) show the color plot of the corresponding velocity field (in the flow direction). Panel~(c) shows 
the time evolution of the total shear stress (blue line) and entropy production (yellow line). Dashed and dotted black lines respectively 
represent the time corresponding to the configuration plotted in panels~(a-d) and~(b-e) respectively. Notice that the total shear stress 
attains approximately null values in the symmetric state while it grows towards larger positive values in the unidirectional regime. 
Conversely the entropy production is larger in correspondence of the superfluidic state. Panel~(f) shows the pdf related to the total 
shear stress. Two peaks at $\sigma_{yz}=0.07$ and $0.7$ are observed corresponding to the two horizontal lines in panel~(c). The most stable 
state is the superfluidic one, compatibly with the fact that it is the state which maximizes the entropy production (yellow line in panel~(c)).}
	\label{fig3}
\end{figure}

Interestingly, large shear stress is a fundamental ingblackient to stabilize the asymmetry. Indeed, at less intense values of external forcing ($\dot{\gamma}<4.0 \times 10^{-4}$) and activity ($ 5 \times 10^{-3} < \zeta < 7.5 \times 10^{-3}$) the asymmetric state is unstable and the  observed dynamics is intermittent with the system jumping from the asymmetric state (see the panel~(a) of Fig.~\ref{fig3}) to a superfluidic regime (panel~(b)), flowing at almost null effective viscosity ($\eta_{eff}=\langle\sigma_{yz} \rangle/\dot\gamma$). The former is analogue to the unidirectional state previously analyzed and it is characterized by the formation of an active layer at just one of the two walls, where the polarization field is oppositely oriented with respect to the imposed velocity (see the enlargement in the inset in panel~(a)). The resulting active force produces a consistent slow-down in proximity of the upper wall, causing the system to  flow mostly leftwards (see panel~(d) of Fig.~\ref{fig3} and dark blue velocity profile in the inset of Fig.~\ref{fig1}(b)).

\begin{figure}[t]
	\centerline{\includegraphics[width=1.0\columnwidth]{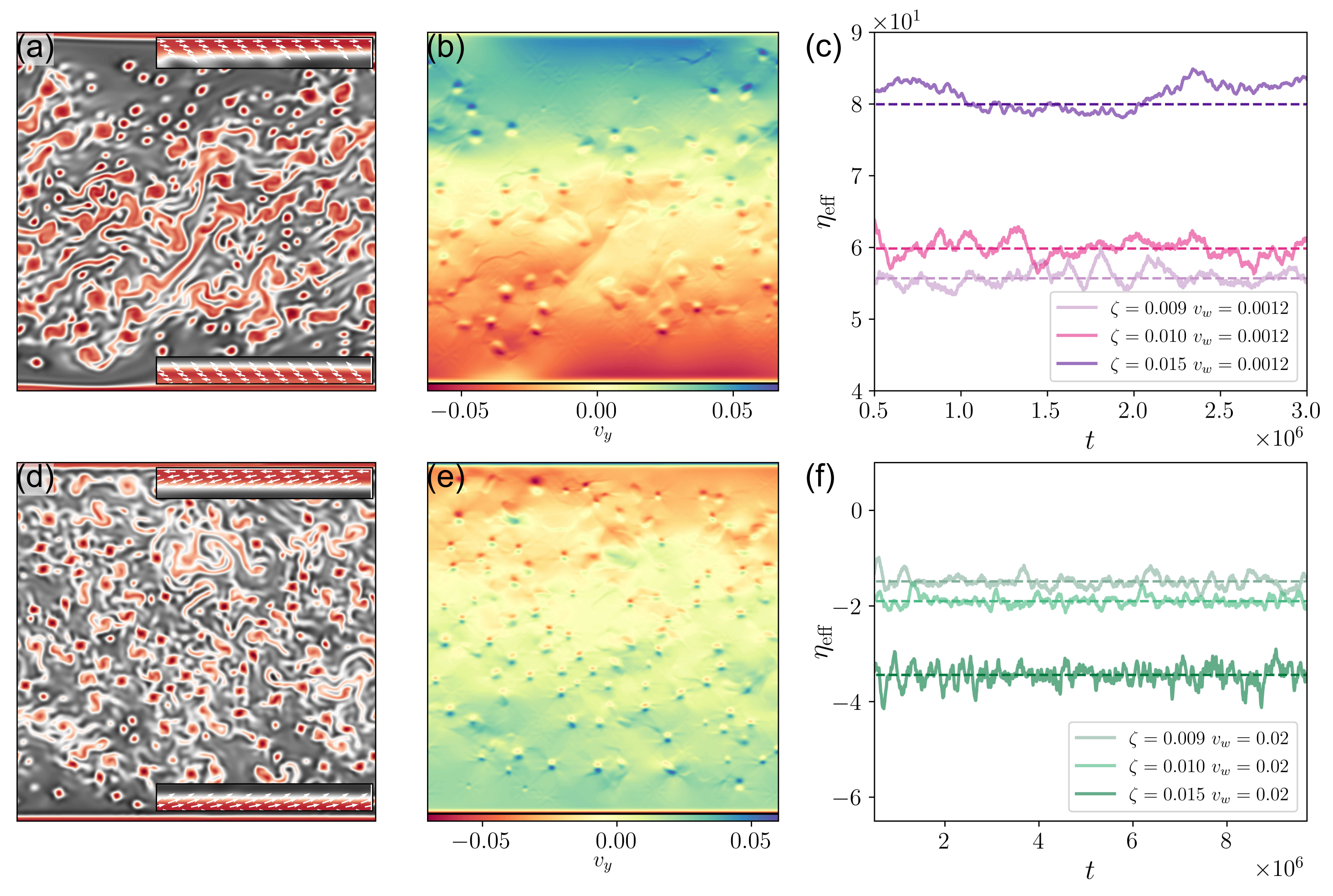}}
	\caption{\textbf{Shear thickening and negative effective viscosity} Panels~(a) and~(d) show the color plot of the concentration 
field $\phi$ at $\zeta=9 \times 10^{-3}, 10^{-2}$ and $\dot{\gamma}=9.3 \times 10^{-6},7.8\times 10^{-5}$, in the PD and NI regions, respectively. 
In the PD case (panel (a)) the polarization field is oriented parallel to the wall velocity (inset of panel (a)) giving rise to an active 
force that enhances the imposed flow (color plot of the velocity component $v_y$ in panel (b)). 
In the NI regime (panel (d)), the polarization field at the boundaries is inverted with 
respect to the wall velocity (insets of panel (d)). In this case the active force is able to invert the flow (color plot of $v_y$ in panel (e)), 
so that the shear stress $\sigma_{yz}$ attains negative values (panel (f) shows the effective viscosity $\eta_{eff}=\sigma_{yz}/\dot\gamma$). Both in the  PD and NI regimes the effective viscosity increases 
(in modulus) with $\zeta$ (panels (c) and (f)). }\label{fig4}
\end{figure}

Importantly, in the limit of weak external forcing, activity is able of counteracting the imposed flow.
Indeed, active features migrate in the channel and eventually adhere to the polarization-free wall (the bottom one for the case in Fig.~\ref{fig3}(a)) and eventually grow forming an active layer thus recovering top-bottom symmetry, as shown in panel~(b). The polarization field aligns in the opposite direction with respect to the imposed velocity as shown by the two insets at the bottom and top right of the panel.
This has the important effect of blackucing the flow intensity at both layers so that the velocity in the bulk of the channel is almost uniform and drastically lower than the imposed one (see the corresponding velocity color plot in Fig.~\ref{fig3}(e) and the related velocity profile plotted in light blue in the inset of Fig.~\ref{fig1}(b)).
Such symmetric configuration is not stable either, as the layers get easily disrupted and eventually vanish, bringing the system back to the unidirectional regime. This multistable dynamics is unambiguously reflected by the time evolution of the shear stress (shown in Fig.~\ref{fig3}(c)) which jumps between  positive and approximately null values corresponding to the asymmetric and symmetric states, respectively.
Interestingly, the probability distribution functions (pdf) of the total shear stress $\sigma_{yz}$ in panel~(f) suggests that symmetric states with vanishing viscosity live longer than the other ones. This behavior can be explained in terms of the rate of entropy production $\Sigma$ which for our system can be written as~\cite{negro2019}
\begin{equation}
\Sigma = 2 \eta_0 \tilde{D} : \tilde{D} + \dfrac{1}{\Gamma} \mathbf{h} \cdot \mathbf{h} + M (\nabla \phi)^2.
\end{equation}
We observe that $\Sigma$ is systematically larger in those time windows where the system sets in the superfluidic state while it drops towards smaller values when the observed state is unidirectional.
Such behavior was already observed in~\cite{negro2019} where it was put forward the hypothesis that a maximum entropy production principle (\emph{MaxEPP}) may hold in selecting the most stable states in multistable active systems.

\section{Large activity: shear thickening and negative effective viscosity}
\label{sec:large}

As activity is increased over $\zeta>0.8 \times 10^{-2}$ internal forcing due to the active injection has a stabilizing effect on the behavior of the system and asymmetric configurations are not observed anymore regardless of the intensity of the shear rate (see Fig.~\ref{fig1}(a)).
The resulting dynamics gives rise to stable regimes characterized either by shear thickening (PD regime) or negative viscosity states (NI regime).

The former occurs at $\dot{\gamma}<1.0\times 10^{-4}$. In this case, the bulk of the channel is populated by rotating droplets while two active layers adhere to the walls (panel~(a) of Fig.~\ref{fig4}). Herein the polarization field is oriented parallel to the wall velocity (insets of Fig.~\ref{fig4}(a)) so that the resulting active force sustains the external forcing, giving rise to enhanced flow profiles 
(see purple line in Fig.~\ref{fig1}(a) and the color plot of $v_y$ in panel~(b) of Fig.~\ref{fig4}). 
However, the region of stability of such regime is limited to low values of shear rate. In this case the effective viscosity
increases with activity as shown in  Fig.~\ref{fig4} (c). 

In the case of larger shear rates the morphology remains basically unalteblack (panel~(d) of Fig.~\ref{fig4}). 
However, the behavior of the polarization field at the boundaries is inverted, \emph{i.e.} $\mathbf{P}$ is oppositely oriented with respect to the wall velocity, as it can be appreciated looking at the polarization field close to the boundaries in the insets of Fig.~\ref{fig4}(c). Mechanistically, the dynamical effect on the flow structure is analogous to the  superfluidic regime, with the active force opposing to the external one and producing a blackuction of the velocity in proximity of the walls.
However now, the active force is considerably stronger and it is able to invert the flow in the thin active layer and boost the fluid in the opposite direction rather than along
the imposed one (see the color plot of $v_y$ in Fig. 4 (e)). This gives rise to the inverted green velocity profile in Fig.~\ref{fig1}(b).
Therefore, the effective viscosity measublack at late times  attains negative values (panel~(f) of Fig.~\ref{fig4} shows the time evolution of the effective viscosity $\eta_{eff}$ for three values of activity) 
due to the counteracting response of the active fluid to external forcing. 

Importantly, both in the case of PD and of NI regimes, the shear stress $\langle \sigma_{yz} \rangle$ increases (in modulus) with $\zeta$, thus leading to a corresponding increment of $|\eta_{eff}|$ as shown in panels (c) and (f) of Fig.~\ref{fig4}.
This suggests that activity may either induce shear thickening in the PD regime, or shear thinning at large shear rates, producing states 
which flow with effective viscosity that attains more and more negative values at larger values of $\zeta$.

\section{Conclusions and discussion}
In this paper we carried out a systematic numerical investigation of a confined $2d$ active polar emulsion. We have shown that the mutual effect of external forcing and active energy injection allows for selecting and stabilizing different rheological regimes. 

In particular, in the weak activity limit, the system sets into an emulsion of polar droplets suspended in a passive isotropic background.
At low shear rates the system is hexatically ordeblack with few dislocations that ultimately affect the macroscopic flow which exhibits shear bandings.  \textcolor{black}{By increasing the shear rate, the imposed flow is able to break such ordeblack structure by deforming the droplets and the velocity profiles become linear.}  Increasing the activity over a certain  threshold greatly affects the morphological properties of the system. Droplets in the bulk begin to merge with each other generating large active domains while a  thick  active layer forms 
at the moving boundaries where the polarization field is tangentially anchoblack. 
At intermediate  values  of  activity  and  large  shear  rate,  velocity blackuction  is  observed  at only  one  of  the  two  layers  and the system basically flows  in  just  one  direction (unidirected motion). In  the  same  range  of  activity,  blackucing  the  intensity  of  shear rate,  a  region  of  multistability between different rheological states is encounteblack. In this case, the system undergoes an  intermittent  dynamics, jumping from  an  unidirectional  state   to a superfluidic state flowing at almost null effective viscosity.
In this case the velocity profile shows a drastic blackuction at both walls and exhibits a flat region in the bulk. We characterized this regime looking at the pdf of the shear stress, finding that states with lowest shear stress are the most probable and correspond to the maximum rate of entropy production. As activity is further increased, active injection  gives rise to stable regimes characterized either by shear thickening(PD regime) or negative viscosity states (NI regime) for low and high values of external forcing, respectively.

\bibliographystyle{unsrt}
\bibliography{ref}
\end{document}